# Time-independent theoretical framework for stroboscopic nonlinear dynamics based on time-nonlocal response

Yuhui Zhuang[1†], Jiaxin Li[1†], Haidong Li[1], Siyu Li[1], Xiaobin Peng[1], Juan Wu[2], Jiameng Zhang[1], Mingjing Fan[1], Xiaoqin Huang[1], Yi Hu[1*], and Jingjun Xu[1**]

[1]*Key Laboratory of Weak-Light Nonlinear Photonics, Ministry of Education, School of Physics, Nankai University, Tianjin 300071, China*
[2]*Pengcheng Laboratory, Shenzhen 518052, China*
[3]*jjxu@nankai.edu.cn*
**yihu@nankai.edu.cn*
[†]These authors contribute equally to this work.

**Abstract:** Recent experiments have demonstrated the ability to manipulate nonlinear interactions via time modulation, giving rise to the so-called stroboscopic nonlinearity. To date, however, this phenomenon has not been subjected to a rigorous theoretical analysis. In this work, we clarify the physical mechanism underlying stroboscopic nonlinear dynamics based on time-nonlocal response and establish an effective time-independent model under suitable modulation conditions. The proposed model almost exactly reproduces the full time-dependent dynamics in the quasi-steady state and significantly outperforms empirical descriptions used previously. Our results provide a clear physical picture of stroboscopic nonlinear dynamics, and can be extended to other systems with time-nonlocal response, establishing a general framework for engineering nonlinear interactions through temporal modulation.

## 1. Introduction

Time modulation provides a powerful tool for manipulating physical systems across diverse fields. By introducing time-dependent parameters, such as forces, potentials, couplings, or material properties, one can give a system novel characteristics that are not accessible in its static counterpart. For instance, time modulation can stabilize otherwise unstable states [1], induce parametric instability [2], break time-reversal symmetry [3], open momentum gaps [4], and lead to spontaneous breaking of time-translation symmetry [5]. Owing to these interesting functionalities, time modulation has been extensively applied in both classical and quantum regimes, giving rise to a variety of novel phenomena [6-20], which hold great promise for the development of innovative technologies.

Time modulation has long been used to manipulate photorefractive nonlinear dynamics, including nonstationary holographic recording and optical amplification under alternating electric fields [21-25], as well as quasi-steady-state solitons under square-wave bias [26]. Quite recently, such a scheme was improved to introduce more novel dynamics. Self-focusing and -defocusing nonlinearities, tailored to occur stroboscopically, can interplay, yielding the nonreciprocal light–light interactions that violate the action–reaction principle [27]. This type of nonreciprocity, distinct from that based on time-reversal symmetry breaking, brings about novel and interesting phenomena, such as optical predator–prey dynamics [28] and unusual impulse–momentum relationships [29]. Although an empirical model, synthesized without incorporating time dependence, has been used to describe the stroboscopic nonlinearity and the associated experimental observations, the underlying physics and a complete theoretical framework remain lacking. In particular, it is still unclear how time modulation controls the form of nonlinear interaction and why a model independent of time has the potential to fit experimental realities.

In this work, we rigorously derive a time-independent model to describe the stroboscopic nonlinear dynamics. Starting from the band-transport model that governs the light–matter interactions in photorefractive crystals, we show that, the stroboscopic dynamics eventually reach a quasi-steady state, provided that the stroboscopic period is much shorter than the crystal's dielectric time constant. During the quasi-steady evolution, the stroboscopic dynamics can be approximately described by a time-independent model, where the two inverted nonlinear processes occurring in alternating time intervals effectively coexist simultaneously. Notably, the form of the derived nonlinearity has never been reported before in materials without time modulations. This approximate model shows excellent agreement with the full time-dependent model in capturing stroboscopic nonlinear dynamics and significantly outperforms empirical approaches. Our results clarify rigorously the physical mechanism underlying the stroboscopic nonlinear dynamics and demonstrate the potential to engineer optical nonlinearities unattainable in static systems.

## 2. Theory

We start with the following equation that is derived from the well-known band-transport model in a photorefractive crystal [30-32]:

$$\frac{\partial}{\partial x}\left(\frac{\partial E\left(x,z,t\right)}{\partial t}+Q\left(x,z,t\right)E\left(x,z,t\right)\right)=0. \tag{1}$$

Here $E$ is the summation of an applied electric field (denoted as $E_{ext}$) and a space-charge electric field (denoted as $E_{sc}$), and is thus defined as the total electric field. In Eq. (1), $Q=1+|u|^2$ with $u$ being an optical field, while $t=\tau/\tau_d$ is the normalized time, with $\tau$ being the physical time and $\tau_d$ being dielectric time constant. This time constant is governed by the crystal properties and dark illumination conditions, which are typically controlled by applying a background light source to the crystal. In general, $\tau_d$ ranges from milliseconds to tens of hours [33, 34]. In what follows, we only consider a single transverse direction (i.e., the *x*-axis), leaving the more complex scenario of two transverse dimensions for future investigation. From Eq. (1), one can readily obtain that the term $\frac{\partial E\left(x,z,t\right)}{\partial t}+Q\left(x,z,t\right)E\left(x,z,t\right)$ is *x*-independent. In the areas far away from the optical field (generally having a finite width), corresponding to a condition of $u=0$, the total electric field should be equal to the applied electric field, which is formulated as:

$$\frac{\partial E}{\partial t}+QE=\left(\frac{\partial}{\partial t}+1\right)\lim_{x\to\pm\infty}E=\left(\frac{\partial}{\partial t}+1\right)E_{ext}. \tag{2}$$

With initial conditions of $E\left(x,z,t<0\right)=0$ and $E_{ext}(t=0)=E_0$ (where $E_0$ is a constant, representing the initially applied external electric field), one can obtain the spatiotemporal evolution of the total electric field in the crystal by solving Eq. (2):

$$E\left(x,z,t\right)=\left\{\int_0^t\left[\left(\frac{\partial}{\partial t'}+1\right)E_{ext}\left(t'\right)\right]e^{\int_0^{t'}Q\left(x,z,t''\right)dt''}dt'+E_0\right\}e^{-\int_0^{t}Q\left(x,z,t'\right)dt'}. \tag{3}$$

To realize stroboscopic nonlinear dynamics, the applied electric field is adopted as square waves [Fig. 1(a)], which have a form of:

$$E_{ext}(t)=\begin{cases} E_+, & t\in\left(0,\frac{T}{2}\right]+(n-1)T \\ E_-, & t\in\left(\frac{T}{2},T\right]+(n-1)T \end{cases}, \tag{4}$$

where $n=1,2,...$ and $E_+=-E_-=E_0$. Meanwhile, the optical field is set to consist of two stroboscopic beams: during the time of $E_+$ and $E_-$, the optical field is denoted as $\varphi_+(x,z,t)$ and $\varphi_-(x,z,t)$, respectively [Fig. 1(a)]. Therefore, the term $Q(x,z,t)$ in each half of the period has the following form:

$$\begin{cases} Q_+(x,z,t)=1+\left|\varphi_+(x,z,t)\right|^2, & t\in\left(0,\frac{T}{2}\right]+(n-1)T \\ Q_-(x,z,t)=1+\left|\varphi_-(x,z,t)\right|^2, & t\in\left(\frac{T}{2},T\right]+(n-1)T \end{cases}. \tag{5}$$

Equations (3-5) determine the total electric field, which in turn modulates the propagation of the two stroboscopic beams (here set to be extraordinarily polarized) through the Pockels effect:

$$i\frac{\partial\varphi_+(x,z,t)}{\partial z}=-\frac{1}{2n_0k_0}\nabla_\perp^2\varphi_+(x,z,t)-k_0\Delta n(x,z,t)\varphi_+(x,z,t), \tag{6}$$

$$i\frac{\partial\varphi_-(x,z,t)}{\partial z}=-\frac{1}{2n_0k_0}\nabla_\perp^2\varphi_-(x,z,t)-k_0\Delta n(x,z,t)\varphi_-(x,z,t), \tag{7}$$

$$\Delta n(x,z,t)=-\frac{1}{2}n_0^3r_{33}E(x,z,t), \tag{8}$$

where $k_0$ is the wave vector in vacuum, $n_0$ is the unperturbed refractive index of the crystal, $\Delta n$ is a light-induced refractive index change, and $r_{33}$ is an electro-optic coefficient.

At the input (i.e., $z$ = 0), the light intensity distribution $\left|\varphi_+(x,0,t)\right|^2$ and $\left|\varphi_-(x,0,t)\right|^2$ remain time-invariant, and so do the terms $Q_+(x,0,t)$ and $Q_-(x,0,t)$. In this framework, the total electric field can be obtained through calculating Eqs. (3-5). By using $E_{sc}=E-E_{ext}$, one can further derive the space-charge electric field that has the following form in each half period over the interval $(N\text{-}1)T$ to $NT$ ($N$ is an integer):

$$\begin{aligned} E_{sc,+}=&\frac{E_0}{Q_+}\left(1-e^{-Q_+(t-(N-1)T)}\right)+\left(e^{-Q_+(t-(N-1)T)}-1\right)E_0 \\ &+E_0\frac{1-Q_+}{Q_+}\left(1-e^{-Q_+\frac{T}{2}}\right)\frac{1-e^{-(Q_++Q_-)\frac{T}{2}(N-1)}}{1-e^{-(Q_++Q_-)\frac{T}{2}}}e^{-Q_-\frac{T}{2}}e^{-Q_+(t-(N-1)T)} \\ &-E_0\frac{1-Q_-}{Q_-}\left(1-e^{-Q_-\frac{T}{2}}\right)\frac{1-e^{-(Q_++Q_-)\frac{T}{2}(N-1)}}{1-e^{-(Q_++Q_-)\frac{T}{2}}}e^{-Q_+(t-(N-1)T)},t\in\left(NT-T,NT-\frac{1}{2}T\right], \end{aligned} \tag{9}$$

$$
\begin{aligned}
E_{sc,-} = & -\frac{E_0}{Q_-}\left(1-e^{-Q_-\left(t-\left(N-\frac{1}{2}\right)T\right)}\right)-E_0\left(e^{-Q_-\left(t-\left(N-\frac{1}{2}\right)T\right)}-1\right) \\
& +E_0\frac{1-Q_+}{Q_+}\left(1-e^{-Q_+\frac{T}{2}}\right)\frac{1-e^{-(Q_++Q_-)\frac{T}{2}N}}{1-e^{-(Q_++Q_-)\frac{T}{2}}}e^{-Q_-\left(t-\left(N-\frac{1}{2}\right)T\right)} \\
& -E_0\frac{1-Q_-}{Q_-}\left(1-e^{-Q_-\frac{T}{2}}\right)\frac{1-e^{-(Q_++Q_-)\frac{T}{2}(N-1)}}{1-e^{-(Q_++Q_-)\frac{T}{2}}}e^{-Q_+\frac{T}{2}}e^{-Q_-\left(t-\left(N-\frac{1}{2}\right)T\right)}, t\in\left(NT-\frac{1}{2}T, NT\right].
\end{aligned}
\tag{10}
$$

The first two terms in Eqs. (9) and (10) describe the establishment of the space-charge electric field during the last cycle, while the 3rd and 4th terms, decaying exponentially with time, account for the space-charge electric field accumulated before the last cycle.

The time evolution described by Eqs. (9) and (10) is illustrated by a typical example, where we set $\varphi_+ = 3e^{-x^2/w^2}$ and $\varphi_- = e^{-x^2/w^2}$ ($w = 12\ \mu\text{m}$) at $z$ = 0 and choose the normalized stroboscopic period as $T$ = 0.02. Figure 1(b) plots the evolution of the space-charge electric field. Here only the region of $x < 0\ \mu\text{m}$ is shown considering the inversion symmetry about $x$ = 0. The space-charge electric field builds up from zero and eventually reaches a steady state. These dynamics are further verified by the result presented in Fig. 1(c), which is obtained by directly simulating Eq. (2). More details are shown in Fig. 1(d) where the dynamics at $x = 0\ \mu\text{m}$ are plotted. During the evolution, the space-charge electric field also undergoes small oscillations due to the applied square wave electric field; consequently, the final state corresponds to a quasi-steady state.

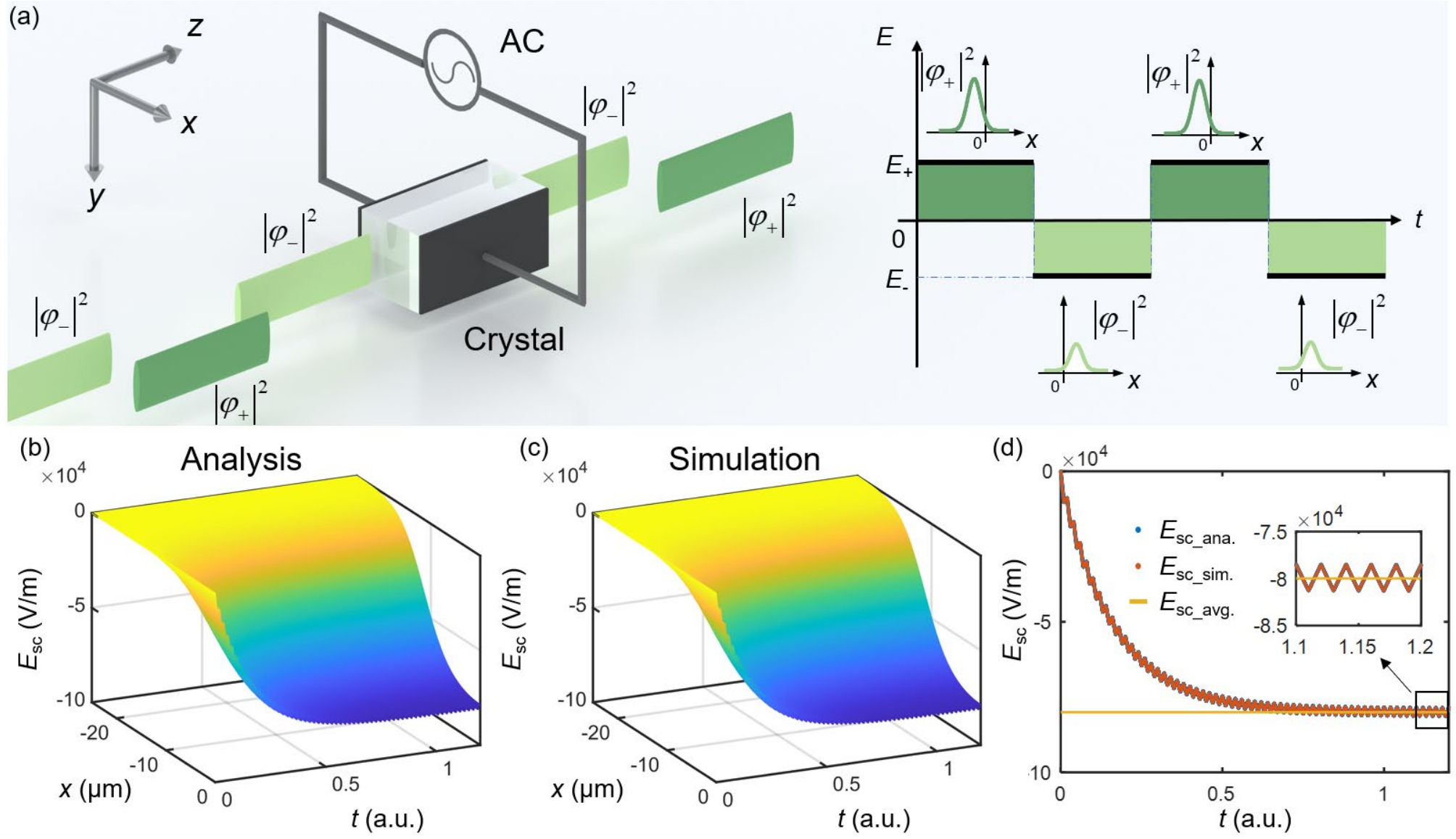

Fig. 1. (a) Schematic setup for realizing stroboscopic nonlinear dynamics: two stroboscopic optical fields (i.e., $\varphi_\pm$) are launched during positive or negative electric field that is externally applied to a photorefractive crystal (see the inset). (b, c) Evolution of the space-charge electric field calculated by Eqs. (9) and (10) (b) and simulated by Eq. (2) (c). (d) shows the dynamics of (b, c) at $x$ = 0: red and blue dots correspond to (b) and (c), respectively, while the yellow solid line denotes the averaged value in the quasi-steady state.

For the propagation cross-section at $z = z_0$ ($z_0 > 0$), the refractive index change for $z < z_0$ varies much with time when the system has not reached a quasi-steady state, and so do the light distributions $\varphi_{\pm}\left(x, z_0, t\right)$. This time-dependent feature makes it impossible to analytically derive the space-charge electric field at $z = z_0$ using the approach deriving Eqs. (9) and (10). However, once a quasi-steady state is built, presumably at $t = t_0$, the light distributions $\varphi_{\pm}\left(x, z_0, t\right)$ can be approximately regarded as unchanged with time. Thus, the time-evolving space-charge electric field for $t > t_0$ can be calculated; its form is similar to Eqs. (9) and (10), except that the time variable needs to be replaced with $t' = t - t_0$ and a term including the space-charge electric field accumulated in the period of $0 < t < t_0$ should be added, i.e., $E_{sc}(x, z_0, t_0) e^{-(Q_+ + Q_-)\frac{T}{2}(N-1)} e^{-Q_+(t' - (N-1)T)}$ and $E_{sc}(x, z_0, t_0) e^{-(Q_+ + Q_-)\frac{T}{2}(N-1)} e^{-Q_+\frac{T}{2}} e^{-Q_-\left(t' - \left(N - \frac{1}{2}\right)T\right)}$ for Eq. (9) and Eq. (10), respectively. The additional term will eventually vanish when sufficient time elapses (i.e., $N \gg 1$), such that the time evolution of the space-charge electric field can still be described by Eqs. (9) and (10). In general, more time is needed to build the quasi-steady state for a longer crystal. Besides, the building time is also determined by the initial distribution of the injected beams. In what follows, we compare the evolution of the space-charge electric field by employing two input conditions with a slight difference. The first involves the injection of two stroboscopic Gaussian beams that are spatially overlapped at the input [Figs. 2(a, b)]; the second is identical, except that a small spacing is introduced between the two beams [Figs. 2(c, d)]. At the input, the building time is nearly the same for both cases, but at the output, it is longer for the latter, indicating that more complex input beam profiles tend to require more time to reach a quasi-steady state.

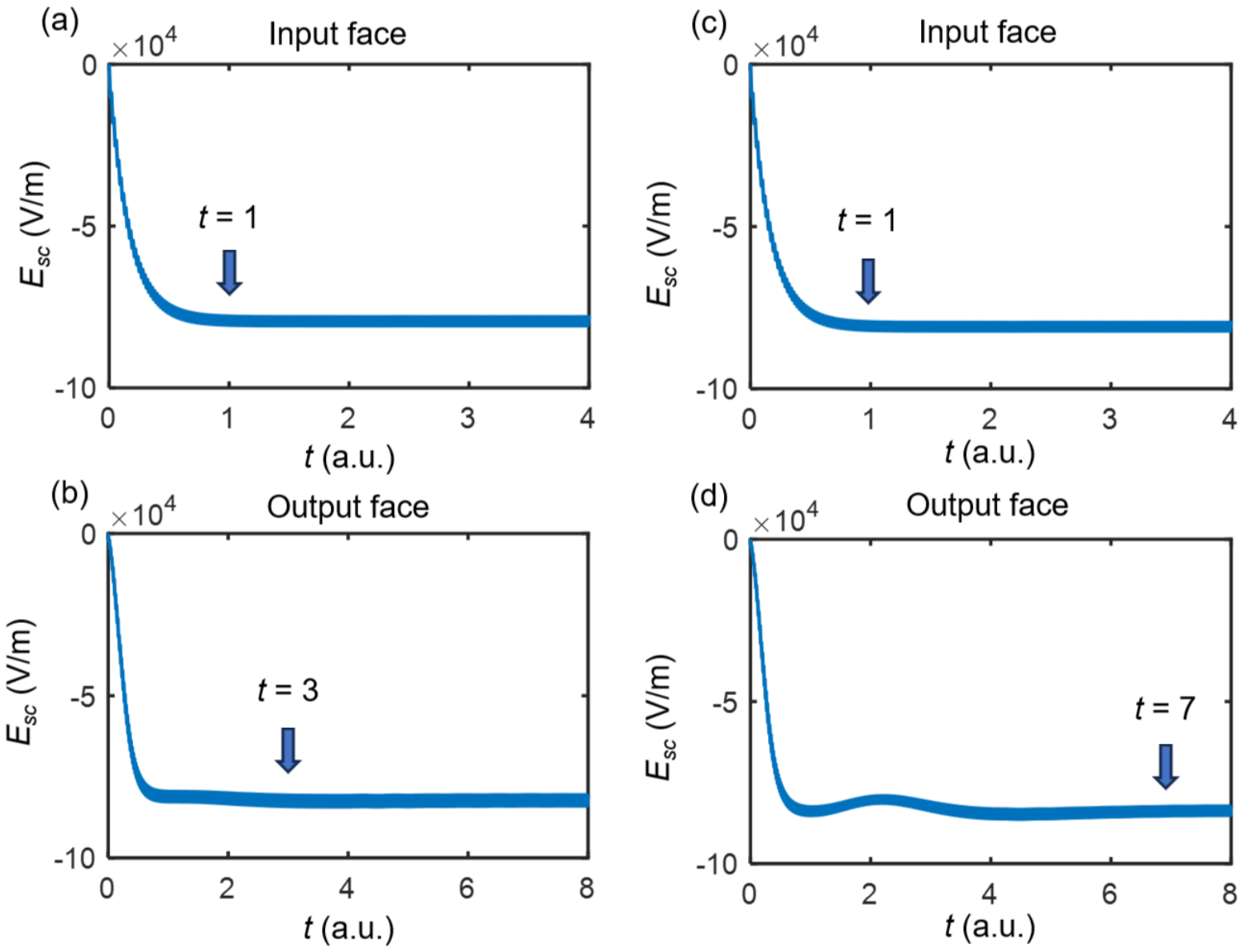

Fig. 2. Evolution of the space-charge electric field (at $x = 0$) at the input (a, c) and output (b, d) by injecting two Gaussian beams with exact overlapping (a, b) or a small spacing (c, d). The arrows mark the time when a quasi-steady state is built.

For the quasi-steady state, the terms in Eqs. (9) and (10) have the following approximation: $\frac{1 - e^{-Q_+\frac{T}{2}}}{1 - e^{-(Q_+ + Q_-)\frac{T}{2}}} \approx \frac{Q_+}{Q_+ + Q_-}$, $\frac{1 - e^{-Q_-\frac{T}{2}}}{1 - e^{-(Q_+ + Q_-)\frac{T}{2}}} \approx \frac{Q_-}{Q_+ + Q_-}$, $1 - e^{-(Q_+ + Q_-)\frac{T}{2}N} \approx 1$, $e^{-Q_{\pm}\frac{T}{2}} \approx 1$, and $e^{-Q_-\left(t-\left(N-\frac{1}{2}\right)T\right)} \approx 1$ . These derivations are obtained by using $N \gg 1$ (corresponding to sufficiently long time) and $\left(Q_+ + Q_-\right)\frac{T}{2} \ll 1$ (satisfied considering the normalized stroboscopic period is quite small relative to 1). Then the space-charge electric field described by Eqs. (9) and (10) is approximated as:

$$E_{sc} = -\frac{\left|\varphi_+\right|^2}{2+\left|\varphi_+\right|^2+\left|\varphi_-\right|^2}E_0 + \frac{\left|\varphi_-\right|^2}{2+\left|\varphi_+\right|^2+\left|\varphi_-\right|^2}E_0, \tag{11}$$

and the refractive index change, formulated by Eq. (8), can be obtained by using $E = E_{sc} + E_{ext}$ . Note that the externally applied electric field is spatially uniform, yielding a uniform refractive index change, which almost does not contribute to the intensity evolution and can be safely ignored. Thus the effective refractive index change has a form of

$$\Delta n = \gamma\frac{\left|\varphi_+\right|^2}{2+\left|\varphi_+\right|^2+\left|\varphi_-\right|^2} - \gamma\frac{\left|\varphi_-\right|^2}{2+\left|\varphi_+\right|^2+\left|\varphi_-\right|^2}, \tag{12}$$

where $\gamma = \frac{1}{2}n_0^3 r_{33} E_0$ . Equation (12) shows a competition between self-focusing (the 1st term) and -defocusing (the 2nd term) nonlinearities. By inserting Eq. (12) into Eq. (6) and Eq. (7), one can build a time-independent model for light propagation under the stroboscopic nonlinearity. In the high-frequency limit of rapidly alternating stroboscopic switching, the time-independent model tends to be more accurate for describing the stroboscopic dynamics.

## 3. Simulation of beam dynamics

In this part, we demonstrate the validity of the time-independent model by studying two reported scenarios. One is associated with the breaking of action–reaction principle in the interaction of two beams [27], and the other is related to vector solitons formed under the stroboscopic nonlinearity [29].

### 3.1 Breaking action–reaction principle in light interactions

Two stroboscopic Gaussian beams are employed here. One of them, synchronized with the positive external electric field, feels a self-focusing nonlinearity, while the other with the negative external electric field undergoes a self-defocusing nonlinearity. The spatial profiles of the former and the latter are $\varphi_+ = Ae^{-\left[(x-d)/w\right]^2}$ and $\varphi_- = Ae^{-(x/w)^2}$ , respectively, where the beam spacing is $d$ = 9.6 μm, and the amplitude and width for both beams are $A$ = 1.4 and $w = 13.6\ \mu\text{m}$ . The propagation dynamics of the two beams, calculated using the time-independent model, are presented in Figs. 3(a, b). $\varphi_+$ induces a waveguide to attract $\varphi_-$ , while the latter induces an anti-waveguide to repel the former. As a result, both beams deflect along the same direction, giving rise to net linear momentum. Such dynamics are caused by the applied electric field, indicating that the obtained linear momentum is strongly correlated with the temporal modulation. Both beams deflect downward owing to the breaking of action–reaction principle. For comparison, the beam dynamics are also simulated via the time-dependent model when the system reaches a quasi-steady state, as shown in Figs. 3(c, d), where the light intensity is averaged over one period. The time-independent results ( $I_\pm$ ) are in good agreement with their time-dependent counterparts ( $I_{t\pm}$ ). Indeed, the two cases match well in detail, as illustrated by the comparison of their intensity profiles in Figs. 3(e-h). Furthermore, we compare the empirical model ( $I_{0\pm}$ ) [27] with the time-dependent model [Figs. 3(i-l)].

Apparent discrepancies are observed in their intensity distributions, which invalidates the precision of the empirical model.

In practice, the crystal may also exhibit an intrinsic beam self-bending effect. Consequently, the beam bending driven by nonreciprocal interactions can be either enhanced or suppressed, depending on whether its direction aligns with or opposes that of the self-bending effect.

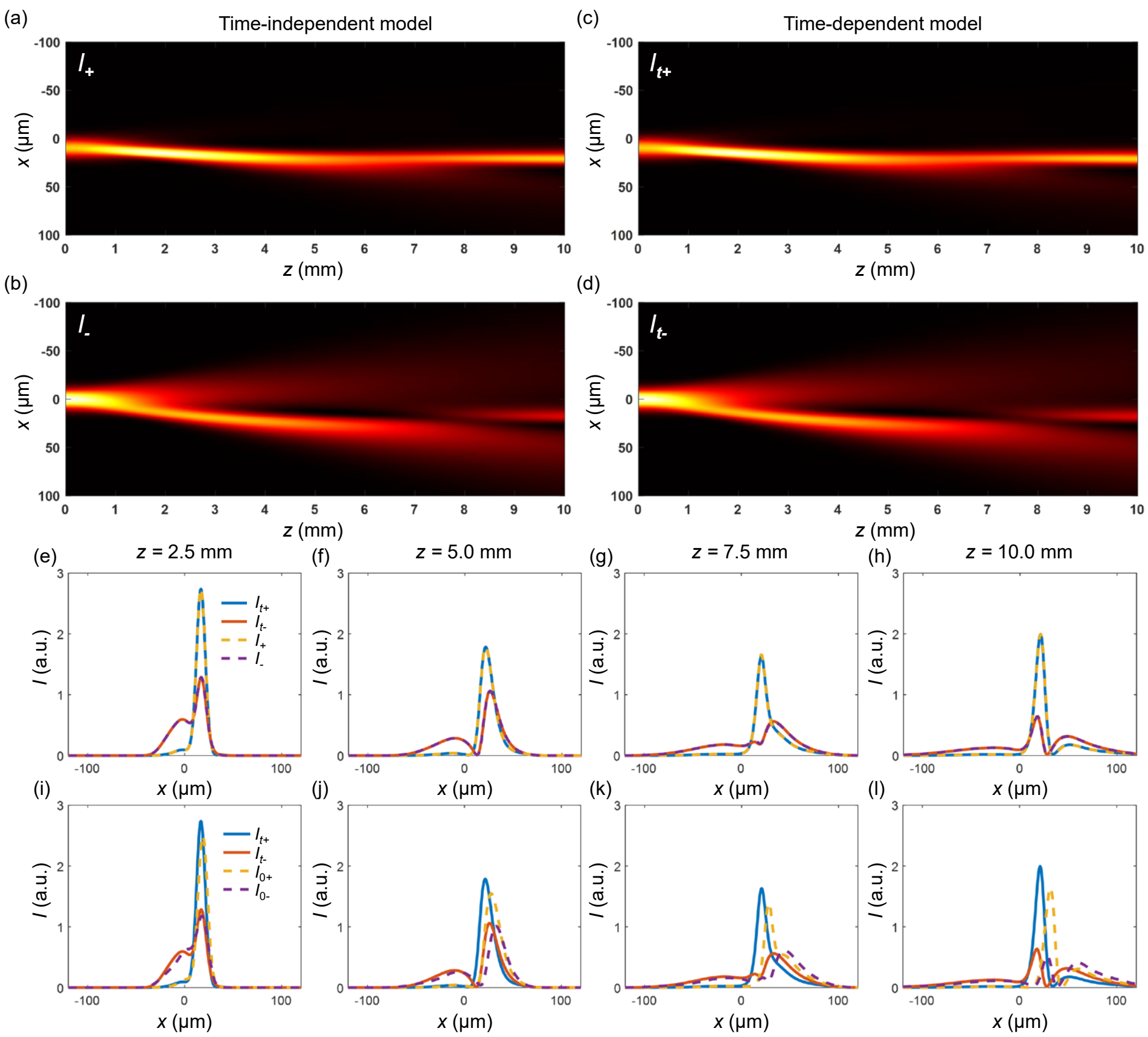

Fig. 3. Simulated interactions of two beams violating action–reaction principle using different models. (a-d) Beam propagation calculated by the time-independent (a, b) and time-dependent (c, d) models. (e-l) Beam intensity profiles at several chosen distances for the beam dynamics based on the time-independent ( $I_\pm$ ), time-dependent ( $I_{t\pm}$ ) and empirical models ( $I_{0\pm}$ ).

## 3.2 Dynamics of a vector soliton

To obtain a vector soliton in the time-independent model, we let $\varphi_+ = u_+(x)\exp(i\beta_1 z)$ and $\varphi_- = u_-(x)\exp(i\beta_2 z)$, where $u_\pm(x)$ are real and $\beta_{1,2}$ are the corresponding eigenvalues. Then Eqs. (6,7,12) are reshaped to:

$$\frac{1}{2n_0k_0}\frac{\partial^2 u_+}{\partial x^2}-\beta_1 u_+ + k_0\gamma\frac{|u_+|^2-|u_-|^2}{2+|u_+|^2+|u_-|^2}u_+ = 0, \tag{13}$$

$$\frac{1}{2n_0k_0}\frac{\partial^2 u_-}{\partial x^2}-\beta_2 u_- + k_0\gamma\frac{|u_+|^2-|u_-|^2}{2+|u_+|^2+|u_-|^2}u_- = 0. \tag{14}$$

For soliton solutions in which fields $u_+$ and $u_-$ are mutually proportional, the eigenvalues, i.e., $\beta_{1,2}$, are equal, so the system can be reduced to a single nonlinear equation. Specifically, letting $u_- = \alpha u_+$ and $\beta_1 = \beta_2 = \beta$, Eqs. (13) and (14) are reduced to a form of

$$\frac{1}{2n_0 k_0}\frac{\partial^2 u_+}{\partial x^2} - \beta u_+ + k_0\gamma\frac{\left(1-\alpha^2\right)}{\left(1+\alpha^2\right)}\frac{\left|u_+\right|^2}{\frac{2}{\left(1+\alpha^2\right)}+\left|u_+\right|^2}u_+ = 0. \tag{15}$$

To guarantee a soliton solution, $\alpha$ is set to be smaller than 1 to ensure that the self-focusing nonlinearity is stronger than the self-defocusing nonlinearity. For illustrative purposes, the soliton solution with a single hump is calculated. Its typical profile is shown in Fig. 4(a). The soliton family is presented in Fig. 4(b), where the eigenvalue $\beta$ is plotted against the amplitude $A_+$ of $u_+$ and the relative amplitude ratio $\alpha$. To provide further details, the soliton width is plotted in Fig. 4(c), defined as the full width at half maximum (FWHM) of the soliton intensity profile. As the amplitude $A_+$ increases, the soliton width initially decreases and then increases. Furthermore, the solitons broaden at larger values of $\alpha$, which strengthens the counteraction between self-focusing and self-defocusing nonlinear effects, ultimately weakening the effective nonlinearity.

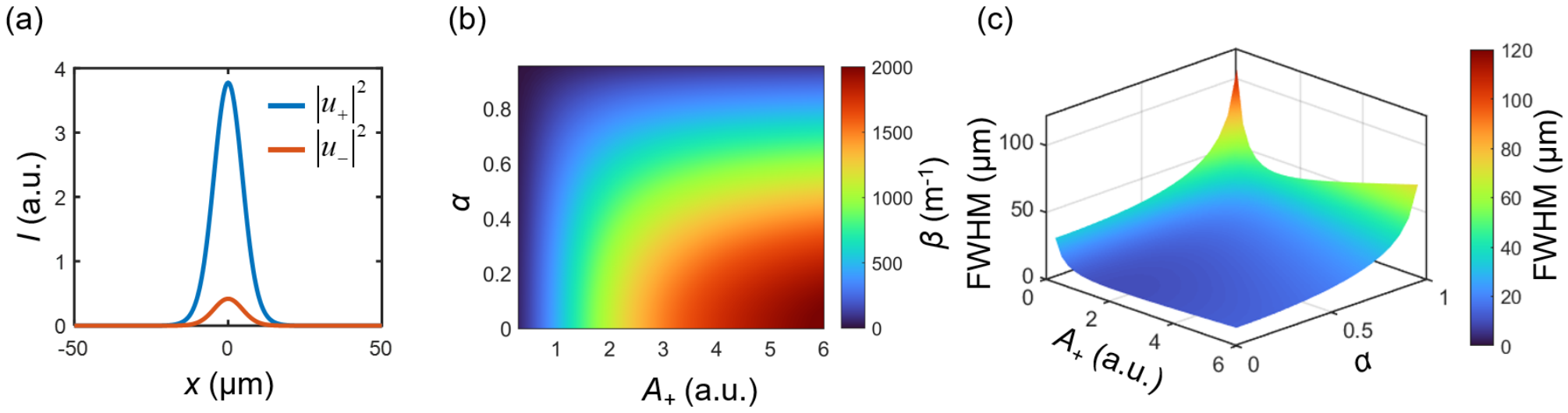

Fig. 4. Soliton solutions obtained with the time-independent model. (a) Typical intensity profile of the soliton solution. (b) Eigenvalue of the soliton as a function of the amplitude $A_+$ and the relative amplitude ratio $\alpha$. (c) FWHM of the solitons in (b).

To compare the soliton dynamics obtained with different models, we define the following parameter:

$$S = \frac{\int I_1(x) I_2(x)\,dx}{\sqrt{\int\left[I_1(x)\right]^2 dx}\sqrt{\int\left[I_2(x)\right]^2 dx}}, \tag{16}$$

which characterizes the similarity of two beam profiles (for instance, $I_1$ and $I_2$) calculated using different models. The parameter $S$ ranges from 0 to 1, with a higher value indicating better similarity. In particular, $S = 1$ denotes an exact overlapping of the two intensity profiles. The comparison between the time-independent and -dependent models is summarized in Fig. 5(a), where the soliton in Fig. 4(a) is used as the input condition. Note that the intensity is averaged in a period in the time-dependent case. The value of $S$ remains nearly 1 along the propagation direction for both components of the vector soliton, indicating that the time-independent model is valid for describing the stroboscopic nonlinearity in the quasi-steady state during the soliton propagation. In contrast, the empirical model shows a distinct deviation when compared to the time-dependent model [Fig. 5(b)] under the input condition of the soliton in Fig. 4(a). Using the parameter $S$, we also compare the three models for calculating the case in Section 3.1, and similar conclusions are obtained [Figs. 5(c, d)]: the time-independent model describes the stroboscopic nonlinear dynamics well.

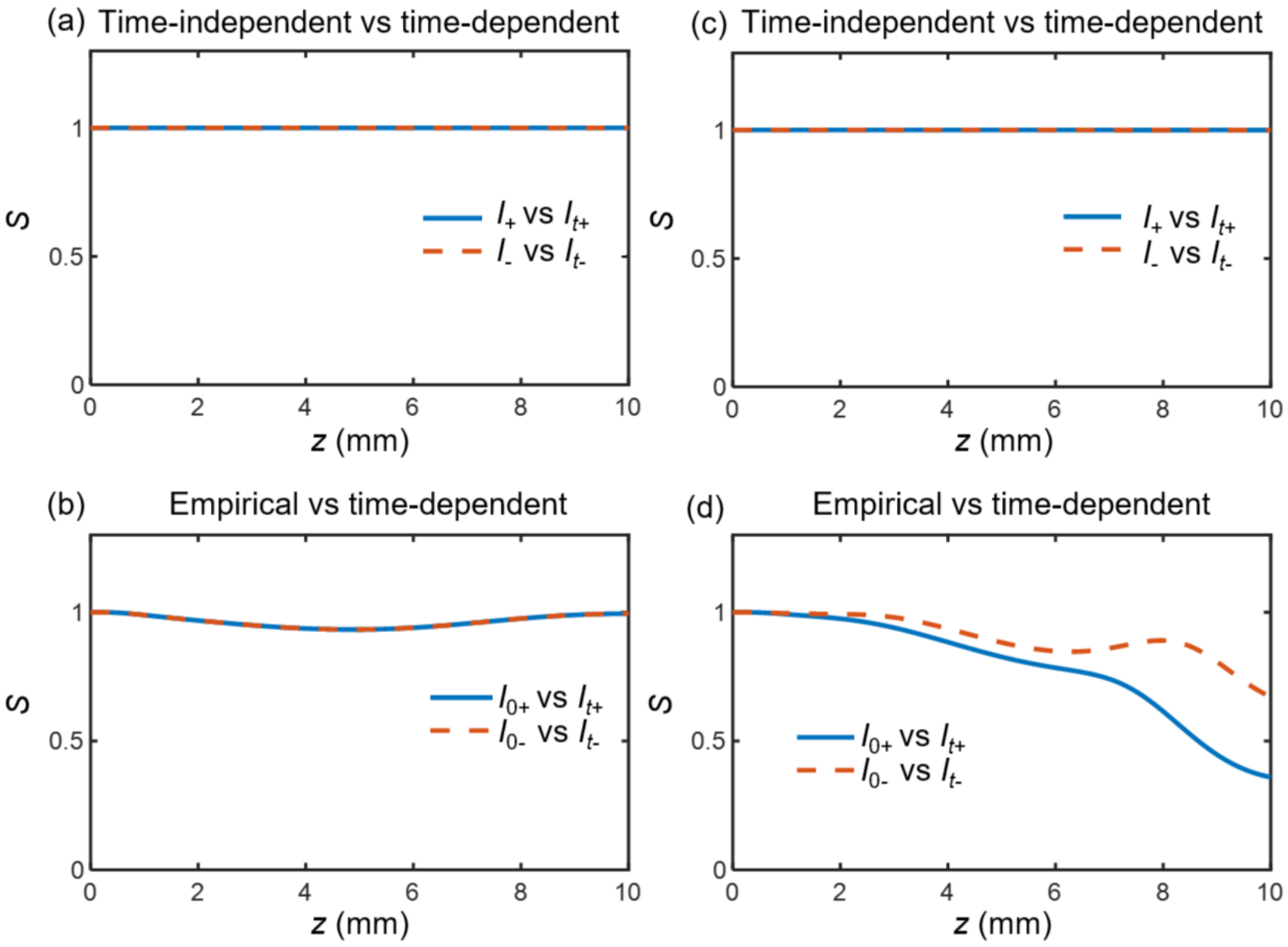

Fig. 5. Comparison of beam dynamics based on different models using the parameter $S$. Top row: time-independent vs time-dependent model. Bottom row: empirical vs time-dependent model. Left and right columns correspond to the dynamics of a vector soliton and two interacting Gaussian beams employed in Section 3.1, respectively.

Furthermore, to compare the dynamics calculated using the time-dependent and time-independent models, we employ more soliton solutions as the input conditions. Without loss of generality, $\alpha$ is set to be 0.5. For each soliton, the similarity of the dynamics based on the two models is first characterized by $S(z)$ in Eq. (16), and then $S(z)$ is averaged over the propagation distance $L$, i.e.,

$$S_{avg} = \frac{1}{L}\int_{0}^{L} S(z)dz. \tag{17}$$

Since the value of $S(z)$ is between 0 and 1, $S_{avg} = 1$ indicates that the two dynamics are identical. The values of $S_{avg}$ for the chosen solitons are summarized in Fig. 6. They are all close to 1, indicating that the time-independent model describes the stroboscopic nonlinear dynamics well.

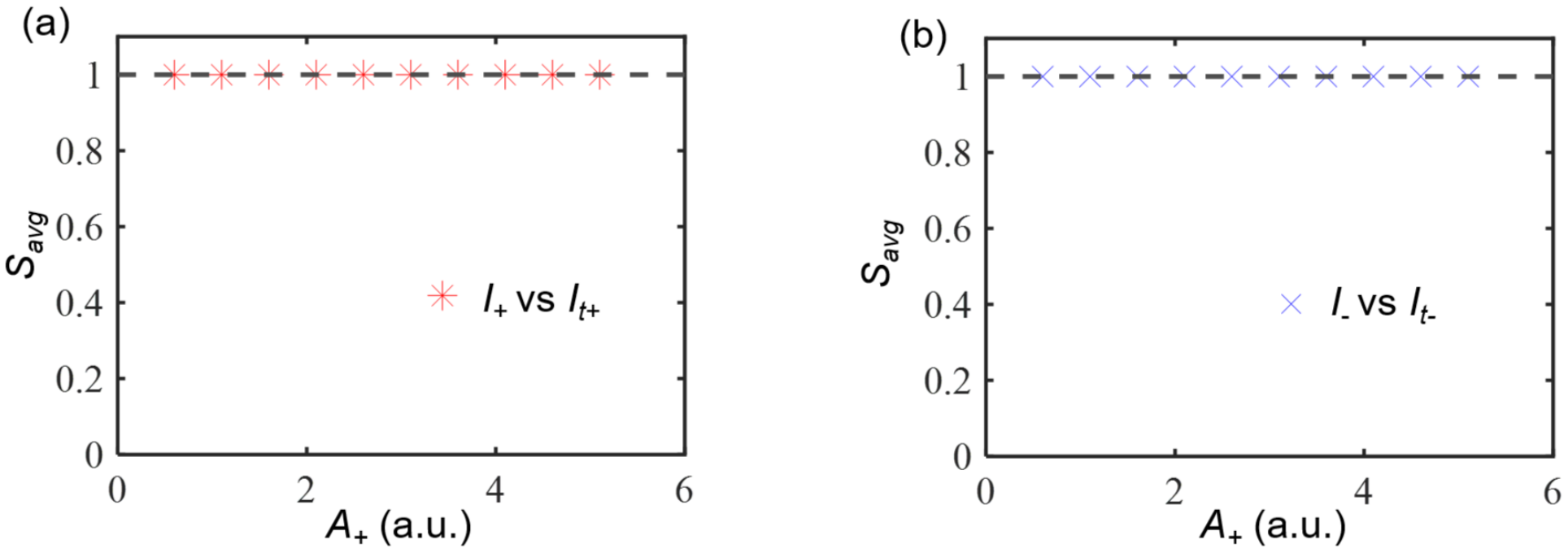

Fig. 6. Similarity between the soliton dynamics calculated with the time-dependent and -independent models. (a) and (b) correspond to the components feeling the self-focusing and -defocusing nonlinearities, respectively. The black dashed lines are at $S_{avg} = 1$ for reference.

## 4. Conclusion

In conclusion, we have revealed the underlying physical mechanism of stroboscopic nonlinear dynamics and established an effective time-independent model to describe them. We find that, when the stroboscopic period is much shorter than the dielectric time constant of the crystal, the dynamics eventually evolve into a quasi-steady state in which opposite nonlinear processes effectively coexist. This coexistence gives rise to an unconventional form of optical nonlinearity that cannot be realized in static systems. Quantitative comparisons indicate that the derived time-independent model describes the quasi-steady state of the time-dependent dynamics well, while generating more accurate calculations than the previously used empirical model. These results provide a clear and rigorous understanding of stroboscopic nonlinear dynamics and highlight time modulation as a powerful strategy for engineering optical nonlinearities.

**Funding.** National Key R&D Program of China (2025YFA1411800); National Natural Science Foundation of China (NSFC) (12250009); 111 Project in China (B23045).

**Disclosures.** The authors declare no conflicts of interest.

**Acknowledgment.** We acknowledge financial support from the National Key R&D Program of China (2025YFA1411800), the National Natural Science Foundation of China (NSFC) (12250009) and the 111 Project in China (B23045).

**Data availability.** Data underlying the results presented in this paper are not publicly available at this time but may be obtained from the authors upon reasonable request.